\documentclass[a4paper,11pt]{article}
\usepackage{jheppub} % for details on the use of the package, please
\usepackage[T1]{fontenc} % if needed
\usepackage{fancyhdr}
\usepackage{graphicx}
\usepackage{amsmath}
\usepackage{amsfonts}
\usepackage{amssymb}
\usepackage{appendix}
\usepackage{color}
\usepackage{cancel}
\usepackage{tensind}
\usepackage{feynmp-auto}
\usepackage{tensor}
\usepackage{dsfont} % double letters
\usepackage{enumitem}
\usepackage{tikz}
\usetikzlibrary{decorations.markings}
\usepackage{subcaption}
\usepackage{physics}
\usepackage{algorithmicx}
\usepackage{amsthm}
\usepackage{leftidx}
\usepackage[export]{adjustbox}
\usepackage{suffix}
\usepackage{mathtools}
\usepackage{tikz-cd}
\usepackage{psfrag}
\usepackage{youngtab}
\usepackage{slashed}
\usepackage{tikz-feynman}
\usepackage{braket}
\usepackage{float}
\usepackage{bm}
\usetikzlibrary{trees}
\usetikzlibrary{decorations.pathmorphing}
\usepackage{verbatim}
\usepackage{tikz-feynman}
\tikzfeynmanset{compat=1.0.0}
\newcommand\mydef{\mathrel{\stackrel{\makebox[0pt]{\mbox{\normalfont\tiny def}}}{=}}}

\title{\boldmath $T\bar{T}$ Deformations in Curved Space from 4D Chern-Simons Theory}

\author{Victor Py}

\affiliation{UC Davis Mathematics and Center for Quantum Mathematics \& Physics,\\One Shields Avenue, Davis, CA 95616, USA}

\emailAdd{vpy@ucdavis.edu}

\abstract{In this paper, we shed new light onto $T\bar{T}$- deformations by engineering them on 2D surface defects, supporting chiral and antichiral CFT's, in a 4D Chern-Simons bulk. This approach is motivated by various connections between $T\bar{T}$-deformations, gravity and Chern-Simons theory and suggests that the formalism developed in this paper and the gravity picture of $T\bar{T}$-deformations should be closely related.}

\begin{document} 
\flushbottom
\maketitle
%\setcounter{tocdepth}{2}
%%%%%%%%%%%%%%
% Your essay starts here %
%%%%%%%%%%%%%%

\section{Introduction}

In earlier papers (\cite{Mazenc:2019cfg,Cardy:2018sdv,Smirnov:2016lqw,Cavaglia:2016oda}), it was shown that the infinitesimal deformation of a CFT by the determinant of its energy-momentum tensor, a ``$T\bar{T}-$ deformation,''  generates a flow in the space of QFTs along the deformation parameter (see \cite{Jiang:2019epa} for a review of $T\bar{T}$-deformations in 2D). This flow follows a hydrodynamic equation (the inviscid Burgers equation) for the energy levels of the deformed QFT \cite{Cavaglia:2016oda}. It has also been shown that these $T\bar{T}$-deformations, widely studied on flat space previously, could be carried out and formalized in curved space as well \cite{Brennan:2020dkw,Dubovsky:2017cnj,Gross:2019ach,Tolley:2019nmm,Mazenc:2019cfg}. Moreover, \cite{Mazenc:2019cfg} have proven that the partition functions of the QFTs along the flow obey a diffusion-like differential equation in the deformation parameter by introducing a kernel relating the partition functions of deformed and undeformed theories. In that paper, the aforementioned kernel was presented as an ansatz that originated in radial wavefunctions in 3D gravity, which had been shown to be related to CFT partition functions. The same year, \cite{Tolley:2019nmm} drew an explicit parallel between $T\bar{T}$-deformations of QFTs and a coupling to massive gravity. This provides some solid motivation as to why gravity is strongly related to $T\bar{T}$-deformations. Furthermore, since gravity is also very closely related to 3D Chern-Simons theory with gauge group $SL_2$ \cite{Witten:1988hc}, Chern-Simons theory seems like a solid ground to build some further intuitive understanding of $T\bar{T}$-deformations.

In this paper, we will introduce a new perspective on $T\bar{T}$-deformations on curved space, based on Chern-Simons theory. We will show that it is possible to couple a 2D CFT that factors into chiral and antichiral sectors to a 4D version of Chern-Simons theory (introduced by Costello et al. \cite{Costello:2017dso}). We will then argue that the coupling to 4D CS can be used to engineer $T\bar{T}$-deformations, at least perturbatively. The basic idea is to look at the coupled theory at low energy and identify an interaction term proportional to $T\bar{T}$ that deforms the effective 2D action. The coefficient in front of $T\bar{T}$ then corresponds to the metric of the space on which the CFT that undergoes the deformation lives. We will use this to show that our approach can be used to engineer $T\bar{T}$-deformed theories in a variety of curved backgrounds, within a common framework.

The paper is organized as follows. In Section \ref{sec::ttbar}, we shall first swiftly go over the principle behind $T\bar{T}$-deformations in order to give the basic definition and notation that we will be using in the following sections. We shall then define a 4D Chern-Simons theory framework following Costello's prescription and insert two 2D chiral and antichiral CFTs as surface defects in this 4D bulk, and give an example of how such a framework has already been used by \cite{Costello:2019tri} to recover deformations similar to $T\bar{T}$. Section \ref{sec::integout} will start with a brief reminder of how the 4D Chern-Simons setup that we have relates to gravity, before getting to the heart of the matter. Coupling the two CFTs through their stress-energy tensor to the gauge field of the bulk theory and integrating out the gauge field then allows us to recover a 2D effective field theory with an interaction term proportional to $T\bar{T}$ whose dependence in 2D spacetime corresponds to a metric term. We will follow this procedure with different gauge groups and prescriptions for our 4D Chern-Simons theory. We will see in Section \ref{sec::TTbarCP1} that using gauge group $SL_2(\mathbb{C})$ and a holomorphic 1-form $dz$ in the definition of 4D CS theory one can recover a $T\bar{T}$-deformation on $\mathbb{CP}^{1}$ or on $AdS_{2}$; while using the same gauge group but the form $dz/z$ in Section \ref{sec::TTbarsausage} will yield a deformed theory on a space endowed with the so-called sausage metric. Finally, a quick look at another gauge group with similar presentation, namely that of the isometries of the plane, will provide us with a way of getting a deformed theory on a space equipped with Witten's ``cigar metric,'' as shown in Section \ref{sec::cigar} 

These results constitute a proof of concept showing that we can interpret some $T\bar{T}$-deformations of theories on curved space as coming from conformal surface defects in a well-tuned 4D gauge theory. It would be interesting in future work to fully determine on which kinds of geometries $T\bar{T}$-deformations can be recovered with this approach.

\newpage
\section{Brief review of $T\bar{T}$ deformations and 4D Chern-Simons Theory}\label{sec::ttbar}
\subsection{$T\bar{T}$ deformation of conformal field theories}

The principle of $T\bar{T}$-deformations (see \cite{Jiang:2019epa} for further references and developments) is quite simple at first order in perturbation theory: given a 2D conformal field theory with stress tensor $T$ in complex coordinates, one can build an irrelevant operator by taking the determinant of the energy momentum tensor:
\vspace{7mm} 
\begin{equation}
    \text{det}\left(\begin{matrix}
    T_{zz} & 0 \\
    0 & T_{\bar{z}\bar{z}}
    \end{matrix}\right)=T_{zz}T_{\bar{z}\bar{z}}\mydef T\bar{T}\ \ ,
\end{equation}

\vspace{4mm}
\noindent and infinitesimally deforming the 2D CFT with this operator:
\begin{equation}
    S[\lambda]=S[0]+\lambda\int_{2D}\sqrt{-g}\,T\bar{T}\ \ ,
\end{equation}
with $\lambda$ a small deformation parameter. It has been shown that such deformations in flat space do not affect integrability (inherited from the seed theory) \cite{Smirnov:2016lqw} and that these deformations can be generalized to curved space \cite{Mazenc:2019cfg}. Of course, determining a non-perturbative formula for the deformed action or the deformed partition function is incomparably more involved and we will not attempt to describe it in this paper.

Previous works \cite{Dubovsky:2018bmo} proved that the energy levels in the spectrum of the deformed theory (when compactifying on a cylinder) satisfy a well-known non-linear differential equation (Bürger's equation) with dependence in the deformation parameter and the radius of the cylinder, while the partition function itself satisfies a close-to-diffusion flow equation (with the deformation parameter playing the role of "time"). Although we will not delve any further into these considerations, we will also mention that \cite{Iliesiu:2020zld,Dubovsky:2017cnj} showed that $T\bar{T}$-deformations can be interpreted - with some relabeling - as a gravitational perturbation of Jackiw-Teitelboim (JT) gravity, and that through AdS/CFT correspondence in the absence of matter fields, a $T\bar{T}$-deformation of the boundary 2D CFT corresponded to a theory of gravity in AdS at a finite cutoff determined by the deformation parameter. Here, however, we will not follow this route, and will endeavor instead to find a simple way of obtaining $T\bar{T}$-deformed theories from higher dimensional gauge theories. 

\subsection{4D Chern-Simons theory}\label{sec::4DCST}

4D Chern-Simons is a partially holomorphic extension of 3D Chern-Simons introduced by Costello \cite{Costello:2017dso}. It makes sense in particular on spaces of the form $\Sigma\times \mathcal{C}$, with $\Sigma$ a 2-dimensional oriented surface, and $\mathcal{C}$ a complex manifold endowed with a meromorphic one-form. In this paper, we will choose $\Sigma$ to be various spaces (like $\mathbb{CP}^1$) parametrized with a complex variable for convenience (we will not be explicitly using any complex structure on it), and we will use two different $\mathcal{C}$, namely $\mathbb{C}_{z}$ (where the subscript refers to the complex coordinate we put on that space) endowed with the 1-form $dz$ and $\mathbb{C}^{\times}_{z}$ endowed with the 1-form $dz/z$. The gauge part of the action reads:

\begin{equation}
    S=\int_{\Sigma\times\mathcal{C}}\omega\wedge CS(A)\ \ ,
\end{equation}
where the $\omega$ is the aforementioned 1-form on $\mathcal{C}$, and where $CS(A)$ is the standard Chern-Simons 3-form:
\begin{equation}
    CS(A)=\Tr{A\wedge \text{d}A + \frac{2}{3}A\wedge A\wedge A}\ \ ,
\end{equation}

\noindent where the trace is over the fundamental representation of the gauge group.Moreover, it is worth mentioning that since CS theory is topological in the first place, this 4D version of Chern-Simons depends topologically on $\Sigma$ (and holomorphically on $\mathcal{C}$).

Taking the slightly easier example of $\mathcal{C}=\mathbb{C}_z$, the action reads
\begin{equation}
    S=\int_{\mathbb{C}_z\times\mathbb{C}_w}\text{d}z\wedge \Tr{A\wedge \text{d}A + \frac{2}{3} A\wedge A\wedge A}\ \ ,
\end{equation}
and it is useful to note an extra gauge freedom (on top of the usual gauge symmetry associated to Chern-Simons theory) coming from the $dz\wedge$. The usual gauge symmetry is captured by the transformation

\begin{equation}
    A\to A+d_{A}\chi\ \ 
\end{equation}
for $\chi$ an infinitesimal parameter, where we define $d_{A}$ to be
\begin{equation}\label{eq::gaugetsfo}
    d_{A}\chi \mydef d\chi + [A,\chi]\ \ ,
\end{equation}
while the outstanding one coming from wedging with the holomorphic one-form reads
\begin{equation}
    A_z\to A_z + \phi(z,\bar{z})\ \ 
\end{equation}
for any function $\phi$ of $z$ and $\bar{z}$. The latter equation makes the $A_z$ component of the action irrelevant, as the action is then invariant under any shift of that component (including by negative itself) 
. 

Thus, we might as well fix the gauge by choosing $A_z=0$, making our 1-form $A$ effectively read: 
\begin{equation}
    A=A_{\bar{z}}\text{d}\bar{z}+A_w \text{d}w+A_{\bar{w}}\text{d}\bar{w}\ \ .
\end{equation}
Moreover, we can further gauge fix the usual symmetry by choosing a holomorphic gauge:
\begin{equation}
\partial_{z}A_{\bar{z}}=0\ \ ,
\end{equation}
which happens to be related to the Lorentz gauge
\begin{equation}
\partial_{z}A_{\bar{z}}+ \partial_{w}A_{\bar{w}}+ \partial_{\bar{w}}A_w=0
\end{equation}
by using the topological character of Chern-Simons theory to infinitely rescale $\mathbb{C}_w$.
If we now add to this the asymptotic boundary condition
\begin{equation}
    \lim_{|z|\to \infty}A_{\mu}(z)= 0 \ \forall \mu\ \ ,
\end{equation}
we get 
\begin{equation}
    A_{\bar{z}}=0\ \ ,
\end{equation}
leaving only two effective components to our gauge field:
\begin{equation}
    A_{GaugeFixed}=A_w \text{d}w + A_{\bar{w}}\text{d}\bar{w}\ \ .
\end{equation}

\subsection{$J\bar{J}$ deformations}

Recently, Costello and Yamazaki showed that such a 4D Chern-Simons setup could be used to recover a kind of deformations that looks very similar to $T\bar{T}$-deformations, namely, $J\bar{J}$ deformations\cite{Costello:2019tri}. 

The authors introduce what they call ``order defects", which they define to be (surface) defects on which the degrees of freedom are coupled to the bulk theory. Their 2D theory (on the defects) having a global G-symmetry, they choose a coupling involving its current in order for the symmetry to remain manifest. Schematically,
\begin{equation}
    \int_{w\in \Sigma} A\, J\ \ ,
\end{equation}
where $J$ denotes the current of the G-symmetry and $A$ the 4-dimensional gauge field. 

By inserting chiral and anti-chiral order defects at positions $z$ and $z'$ in the holomorphic plane, coupled through the chiral and anti-chiral currents respectively, and by then integrating out the gauge field, they recover a 2D effective theory with an interaction term that reads
\begin{equation}
    \mathcal{L}^{eff}_{int}=\int_{w\in \Sigma} r_{ab}(z-z')J_a (w,\bar{w}) \bar{J}_b (w,\bar{w})\ \ ,
\end{equation}
where $r_{ab}$ is the so-called classical $r$-matrix involved in the scattering process and can just be thought of as the propagator between two components of the 1-form gauge field.

This result inspired us to wonder whether the same can be done for $T\bar{T}$-deformations, so as to recover in a very similar fashion the deformation term (to first order) from a 4D Chern-Simons setup with surface defects involving an $AT$ coupling.

\section{How to get 2D $T\bar{T}$-deformed theories from 4D Chern-Simons}\label{sec::integout}

\subsection{Brief review of the connection to gravity}\label{sec::gravity}

Although this is not the focus of this paper, it is interesting to keep in mind that our 4-dimensional setup is strongly related to theories of gravity. This suggests the existence of a parallel between $T\bar{T}$-deformations of CFT's as studied in our paper and the appearance of the same deformations in a JT gravity context, for instance. Let us therefore give a lightning review on how 4D Chern-Simons relates to gravity.

Starting with the 4-dimensional setup described in the previous Section, the first thing we can do is to compactify the theory on a circle in the $z$ (holomorphic) direction. In other words, we are doing a Kaluza-Klein reduction in the holomorphic direction. If the circle we compactify on is small enough, the fields can be considered constant over it (we throw away the KK modes). Our holomorphic plane becomes (through the usual conformal transformation $z\mydef e^{h+i\theta},\bar{z}
\mydef e^{h-i\theta}$) a cylinder $\mathcal{C}$ with height coordinate $h$ and angle $\theta$. In our gauge ($A_z = 0$) the pullback of the 1-forms and form components by that map yields
\begin{align}
    dz&=(dh+id\theta)e^{h+i\theta}\nonumber \\
    d\bar{z}&=(dh+id\theta)e^{h+i\theta}\nonumber \\
    A_h&=e^{h+i\theta}A_{\bar{z}}\nonumber\ \ ,
\end{align}
and our action becomes 
\begin{align}
    \int_{z,\bar{z}\in \mathbb{C}\, , \, w,\bar{w}\in \Sigma}\frac{dz}{z}\wedge CS\left[A_{\bar{z}},A_w,A_{\bar{w}}\right] = \int_{\theta\in S^{1},h\in \mathbb{R}\, , \, w,\bar{w}\in \Sigma}d\theta\wedge CS\left[A_h,A_w,A_{\bar{w}}\right](h,w,\bar{w})\ \ .
\end{align}

This is a 3D (analytically continued) Chern-Simons for $SL_2(\mathbb{C})$ which, with the choice of a contour, becomes a Chern-Simons theory for $SL_2(\mathbb{R})$. The latter theory is related to (half of) 3D gravity using the standard identification of Chern-Simons gauge field with vielbein and spin connection \cite{Witten:1988hc}.

We can now go one step further and do yet another compactification, this time over a circle in the topological direction. This yields a 2D $BF$ theory, where $F$ is the field strength of the gauge field and where the auxiliary field $B$ is its monodromy around the circle over which we just compactified. It is then well known that 2D JT gravity and BF theory are equivalent \cite{Isler:1989hq,Chamseddine:1989yz}.

Thus, from 4D Chern-Simons theory, one can recover 2D JT gravity using two Kaluza-Klein compactifications. It would be interesting to use this to explicitly relate $T\bar{T}$ deformations in all these formalisms. In this paper, however, we will focus on establishing how to derive $T\bar{T}$-deformations of 2D CFT's using 4D Chern-Simons.

\subsection{Coupling the CFT to the bulk Chern-Simons theory}
After having introduced the 4D version of Chern-Simons theory, we are now ready to add to the mix the 2D CFT of interest. The incorporation of a 2D (chiral + antichiral) CFT will be carried out as follows: at a given point $z_0$ in $\mathcal{C}$, we insert a free 2D chiral CFT, and at a point $z_1$ we insert its free antichiral counterpart. Finally, we couple both of them to the gauge field $A$ (from Chern-Simons) through their stress tensor, as can be expected in order to recover $T\bar{T}$-deformations. As the coupling process is a little complex, we will first write the coupling terms and then explain why such coupling makes physical sense and yields a gauge-invariant action. To perform it, we need a gauge group generated by three elements (which we call $e,f$ and $h$). The interaction part of the action will read:

\begin{equation}
    S_{int}\mydef\int_{\Sigma,\{z=z_0\} \in \mathcal{C}}\left(A^{e}_{\bar{w}}\mathcal{T}^{f}+A^{f}_{\bar{w}}\mathcal{T}^{e}+A^{h}_{\bar{w}}\mathcal{T}^{h}\right)+\int_{\Sigma,\{z=z_1\} \in \mathcal{C}}\left(A^{e}_{w}\mathcal{\bar{T}}^{f}+A^{f}_{w}\mathcal{\bar{T}}^{e}+A^{h}_{w}\mathcal{\bar{T}}^{h}\right)\ \ ,
\end{equation}
with $\mathcal{T}^{a}=c_{a}w^{p_a}T\mydef \mathcal{V}^{a}T$, and  its antichiral counterpart $\bar{\mathcal{T}}^{a}=c'_{a}\bar{w}^{p'_a}T\mydef \bar{\mathcal{V}}^{a}T$. Here $c,c'$ are complex coefficients, $p,p'$ are natural numbers, and all depend on the choice of gauge group. We want to use the property that $T$ (resp. $\bar{T}$) acts as an operator as $\partial_w$ (resp. $\partial_{\bar{w}}$) to build another copy of the Lie algebra (typically resembling a Virasoro algebra) generated by the $\mathcal{T}^{a}$, such that $e^{-S_{int}}$ stays invariant under gauge transformations.

\subsection{Explicit proof of gauge-invariance in the $SL_2$ case}
Let us work out the gauge-invariance for the case where the gauge group is $SL_2$. By construction, the 4D CS action is invariant under the chosen gauge group's transformations, but the coupling terms themselves are not classically gauge invariant. This is however not a problem as long as the variation of $S_{int}$ is equivalent to zero as an operator. To engineer some $\mathfrak{sl}_2$ gauge invariance, as just mentioned, we use the fact that integrating the stress tensor on a closed contour around an operator acts like a derivative with respect to the complex variable. This allows us to construct the following (Virasoro-like) $\mathfrak{sl}_2$ generators from the stress tensor: $\mathcal{T}^{a}\mydef \mathcal{V}^{a}T$ (with $a\in\left\{e,f,h\right\}$) with

\begin{align}\label{eq::virasorocoeffsCP1}
    \mathcal{V}^{e}(w)&=1\nonumber \\
    \mathcal{V}^{f}(w)&=-w^2 \\
    \mathcal{V}^{h}(w)&=-2w\nonumber \ \ ,
\end{align}
and for the anti-holomorphic counterparts:
\begin{align}\label{eq::virasorocoeffsCP1c}
    \bar{\mathcal{V}}^{e}(\bar{w})&=\bar{w}^2\nonumber \\
    \bar{\mathcal{V}}^{f}(\bar{w})&=-1 \\
    \bar{\mathcal{V}}^{h}(\bar{w})&=2\bar{w}\nonumber \ \ .
\end{align}
Note that we could have taken the same expression for the holomorphic and antiholomorphic vertices in principle, but this prescription will yield more interesting results. 

Let us now explicitly check this invariance for the chiral part of the action (the antichiral part will behave in the exact same way), and along only one "direction", say `$e$' (such that $\chi(w,\bar{w})\equiv \chi_{e}(w,\bar{w})\,e$) - since everything is linear, it suffices to check invariance along each generator independently. Note that we shall henceforth use interchangeably $\bar{A}\mydef A_{\bar{w}}$ and $A\mydef A_{w}$ for practical reasons.

Consider the resulting transformation of the barred components of the gauge field:
\begin{equation*}
    A_{\bar{w}}^{a}\to A_{\bar{w}}^{a}+\bar{\partial}\chi^{a} + [A_{\bar{w}},\chi]^{a}\ \ .
\end{equation*}
In particular, if $\chi$ is only along the Lie algebra generator $e$:
\begin{align}\label{eq::varact}
    \delta_{\chi^e}A_{\bar{w}}^{e}&=\bar{\partial}\chi+ 2 A_{\bar{w}}^{h} \chi_e \nonumber\\
    \delta_{\chi^e}A_{\bar{w}}^{f}&=0 \\
    \delta_{\chi^e}A_{\bar{w}}^{h}&=- A_{w}^{f}\chi_e \ \ . \nonumber
    \end{align}
This yields:
\begin{equation}
    \delta_{\chi^e}S_{int}=\int \left(\bar{\partial}\chi^e+2\bar{A}^{h}\chi^{e}\right)\mathcal{T}^{f}+0\cdot \mathcal{T}^{h}+\left(-\bar{A}^{f}\chi^{e}\right)\mathcal{T}^{e}\ \ .
\end{equation}
Thus, 
\begin{equation}\label{eq::varexp}
    \delta_{\chi^{e}}\left(e^{-S_{int}}\right)=-\delta_{\chi^{e}}S_{int}+\left(\delta_{\chi^{e}}S_{int}\right)S_{int}+ \dots
\end{equation}
Now, the game is to show that part of the second term in the variation of the exponential cancels out the first one. This is possible because one of the ``sub-terms" of the second term (the one containing $\bar{\partial}\chi^{e}$) is linear in $\bar{A}$ (like the terms we want to cancel out) and quadratic in $\mathcal{T}$, which allows us to use the OPE of the stress tensor 
\begin{equation}\label{eq::opeTT}
    T(w')T(w)\sim \frac{\partial T(w)}{w'-w}+2\frac{T}{(w'-w)^2}+\frac{c/2}{(w'-w)^4}
\end{equation}
to recover the right kind of terms, up to integration by parts. This will show (up to generalization to the next orders in $\bar{A}$) that one term at order N+1 in $\bar{A}$ cancels all the remaining terms at order N in $\bar{A}$, that is, the terms that were not used to cancel out the order N-1. 

Let us work this out. For the sake of simplicity, since $A=A_w$ does not appear in the part of the action we are considering (only $\bar{A}$), we can partly fix the gauge and impose:
\begin{equation}\label{eq::gaugefix}
    A=0 \,\, \implies \partial \chi = 0\ \ ,
\end{equation}
where the implication comes from disallowing gauge transformations to modify $A$. This allows us to effectively set $[A,\bar{A}]=0=\bar{\partial} A$, which yields a modified version of the equations of motion for $A_{\bar{z}}$ (obtained pre-gauge fixing) in our Chern-Simons theory:
\begin{equation}\label{eq::eomazbar}
    \partial \bar{A} + \bar{\partial}A+[A,\bar{A}]=\partial \bar{A} = 0\ \ .
\end{equation}
This equation will be useful later in the proof. 

Combining Equations \ref{eq::varact} and \ref{eq::varexp} we get a quadratic variation that can be decomposed into two kinds of terms. Terms quadratic in $A$, which will be cancelled by the cubic variation, and terms in $\bar{\partial}\chi^{e}$ that read: 
\begin{equation}\label{eq::proofhalf}
    \int_{w'}\int_{w}\bar{\partial}\chi^{e}(w')\,\mathcal{T}^{f}(w')\left(A^{e}\mathcal{T}^{f}+A^{h}\mathcal{T}^{h}+A^{f}\mathcal{T}^{e}\right)(w)\ \ .
\end{equation}
An integration by part of the antiholomorphic derivative term will pick out the residue of the $\chi\mathcal{T}\mathcal{T}$ OPE and reduce remove one integral sign by virtue of the complex analysis lemma
\begin{equation}
    \partial\frac{1}{\bar{w}}=\delta^{2}(w)\ \ .
\end{equation}
Using Equation \ref{eq::opeTT}, and recalling that our gauge fixing ensures that $\partial \chi^{e}=0$, we can deduce the residues of the $\chi^{e}\mathcal{T}^{f}(w')\mathcal{T}^{a}(w)$ OPEs:
\begin{align}\label{eq::opesttchi}
    Res\left(\chi^{e}(w')\mathcal{T}^{f}(w')\mathcal{T}^{e}(w)\right)&=-(w^2\partial T+4wT)\chi^{e}(w)\\
    Res\left(\chi^{e}(w')\mathcal{T}^{f}(w')\mathcal{T}^{h}(w)\right)&=(2w^3\partial T+8w^2 T)\chi^{e}(w)\\
    Res\left(\chi^{e}(w')\mathcal{T}^{f}(w')\mathcal{T}^{f}(w)\right)&=(w^4\partial T+4w^3 T)\chi^{e}(w)\ \ .
\end{align}
Plugging the equations in \ref{eq::opesttchi} back into \ref{eq::proofhalf} and integrating by part the $\partial T$ dependences using Equations \ref{eq::eomazbar} and \ref{eq::gaugefix} , we get from the originally $\bar{\partial}\chi^{e}$ term:
\begin{align}\label{eq::proof}
    &\,\,\chi^{e}T\left(\bar{A}^{f}\left(-\partial (w^2)T+4w\right)+\bar{A}^{h}\left(-\partial (-2w^3)-8w^2\right)+\bar{A}^{e}\left(-\partial (-w^4)-4w^3\right)\right)\nonumber \\
    &=\chi^{e}\left(2wT\bar{A}^{f}-2w^{2}T\bar{A}^{h}+ 0T \bar{A}^{e}\right)\nonumber \\
    &=\chi^{e}\left(-\bar{A}^{f}\mathcal{T}^{h}+2\bar{A}^{h}\mathcal{T}^{f}\right)\nonumber\\
    &=\delta_{\chi^{e}}S_{int}-\int \bar{\partial}\chi^{e}\mathcal{T}^{f}\nonumber \\
    &=\delta_{\chi^{e}}S_{int}\ \ ,
\end{align}
where we used the fact that $\mathcal{T}^{f}$ is holomorphic to get the last line by integration by part. 

This final expression, coming from the $\bar{\partial}\chi^{e}$ part of the second term in \ref{eq::varexp}, indeed cancels out the first (linear) term of the variation in \ref{eq::varexp}. Similarly, all remaining (non $\bar{\partial}\chi$) components of the second term in \ref{eq::varexp} will be cancelled out by the $\bar{\partial}\chi$ part of the third term, and so on. Let us show that explicitly.

So far, what we have done can be summed up in the following schematic way. For a variation of the action looking like 
\begin{equation}\label{eq::delS}
    \delta S = D + B\ \ ,
\end{equation}
where $D$ (seen as an operator insertion at all possible $w$'s) is the integral term containing $\bar{\partial}\chi$ and $B$ (operator insertion as well) is the rest of the variation, we have the following: 
\begin{equation}\label{eq::proofuseful}
    D\cdot S = B\ \ ,
\end{equation}
where the multiplication here denotes the product of operators (carried out later using OPE's between closest insertions). Thus, 
\begin{equation}
    -\delta S + \delta S \cdot S = -D - B + D\cdot S + B\cdot S = B\cdot S\ \ ,
\end{equation}
where we used Equation \ref{eq::proofuseful} and the fact that $D$ by itself is zero by integration by part (see Equation\ref{eq::proof}). Here, $B\cdot S$ is quadratic in $\bar{A}$ and we have therefore successfully cancelled out all terms linear in $\bar{A}$.

Let us now show the cancellation to all orders. Using equation \ref{eq::delS}, we have that the variation of $e^{-S}$ reads
\begin{equation}\label{eq::variationact}
    \delta\left(e^{-S}\right)=(D+B)\cdot e^{-S}\ \ .
\end{equation}
Moreover, $D$ acts as a derivation on analytic functions. Indeed, since $D$ contains a $\bar{\partial}\chi$ term, as we saw before, it is going to pick out the residues of all the OPEs coming from all operators insertion (picking out the residue of all possible singular points $w_1=w_i$ for $i>1$ and where the $w_i$ (resp. $w_1$) are the coordinate of the $S$ (resp. $D$) operator insertions), and therefore act as a derivation.
This yields
\begin{align}
    D\cdot e^{-S}&=-(D\cdot S)e^{-S}\nonumber \\
    &=-B\cdot e^{-S}\ \ ,
\end{align}
which immediately shows that the variation in Equation \ref{eq::variationact} is zero. This concludes our argument. 

We have therefore established induction and shown the invariance of the action at the quantum level under a gauge transformation along the $e$ direction. The same work can be done for the other generators ($f$ and $h$) of $\mathfrak{sl}_2$ and for the antichiral CFT coupling. Thus, it concludes our proof of the invariance of $e^{-S_{int}}$ under gauge transformations, thereby showing that the proposed coupling between 4D CS theory and 2D CFTs is viable.

\subsection{Integrating out the gauge field}\label{sec::CStoIFT}
Since we are looking to get a 2D $T\bar{T}$-deformed theory, and as the only objects that are not inherently 2D in our theory are the two remaining components of the gauge field, it seems logical at this point to integrate out those components and compute the effective action resulting from this operation. For the sake of simplicity, this section will describe how to integrate out the gauge field (for any chosen gauge group) in the case of a $dw$ holomorphic 1-form. The relevant expressions for the $\frac{dz}{z}$ case (of which the derivation follows the same steps as in this subsection) will be introduced when needed in a later section.

Seeing as there are only two types of vertices of interest: $\propto A_{\bar{w}}\,T$ at $z_0$, and $\propto A_w\, \bar{T}$ at $z_1$ (the vertices we called respectively $A_{\bar{w}}^{a*}\mathcal{T}^{a}$ and $A_{w}^{a}\bar{\mathcal{T}}^{a}$ in the previous subsection):

\vspace{0.5cm}
\begin{figure}[H]
\begin{center}
\begin{tabular}{ll}
\begin{fmffile}{cubic_vertices}  %creates file named three_body to save your diagram
\begin{fmfgraph*}(100,80)    %size of your diagram you can change it as you wish
\fmfleft{e1}              %indicates the particles which will stand on LHS, for sake of simplicity I denoted them as electrons but you can change them as you like
\fmfright{g}

\fmfv{decor.shape=circle, decor.filled=shaded,decor.size=0.2w,label.dist=0.18w}{e1}%indicates the particles will stand on RHS
\fmflabel{$\mathcal{V}^{a}T$}{e1}         %names which will be shown on your PDF
\fmflabel{$A_{\bar{w}}^{b}$}{g}
\fmf{gluon}{e1,g}       %fermion lines, if you don't want arrows simply write "plain"
%instead of "fermion"
\end{fmfgraph*}
\end{fmffile}\hspace{1.5cm}
&
\begin{fmffile}{three_body2}  %creates file named three_body to save your diagram
\begin{fmfgraph*}(100,80)    %size of your diagram you can change it as you wish
\fmfright{e1}              %indicates the particles which will stand on LHS, for sake of simplicity I denoted them as electrons but you can change them as you like
\fmfleft{g}             %indicates the particles will stand on RHS
\fmfv{decor.shape=circle, decor.filled=shaded,decor.size=0.2w,label.dist=0.18w}{e1}
\fmflabel{$\bar{\mathcal{V}}^{b}\bar{T}$}{e1}         %names which will be shown on your PDF
\fmflabel{$A_{w}^{a}$}{g}
\fmf{gluon}{g,e1}
\end{fmfgraph*}
\end{fmffile}

\end{tabular}
\end{center}
\caption{Only two types of vertices participating in the effective action. $a$ and $b$ are such that $\Tr\left(t^{a}t^{b}\right)\neq 0$ with $\Tr$ the non-degenerate invariant bilinear form of $\mathfrak{sl}_2$ and $t^{a}$ the generators of $\mathfrak{sl}_2$.}
\label{feynman::verticesTheo}
\end{figure}

\noindent we will get as an effective interaction vertex all binary combinations of vertices of different types, multiplied by the propagator between the two, whose color part will just be the quadratic Casimir of the gauge group. 
As to the color-stripped propagator, it is as always the Green's function of the color-stripped kinetic operator in the action. The kinetic term $\text{d}z\wedge A\wedge \text{d}A$ is whittled down to $A_{\bar{w}}\partial_{\bar{z}}A_w\, \Omega$ (where $\Omega$ is the canonical top form), yielding the following differential equation for the propagator: 
\begin{equation}
    \partial_{\bar{z}}G(z-z',w-w')=\delta(z-z')\delta(w-w')\ \ ,
\end{equation}
whose solution is simply
\begin{equation}
    G(z-z',w-w')=\frac{1}{2i\pi (z-z')}\delta(w-w')\ \ .
\end{equation}
Putting color and kinematics together, we finally have
\begin{equation}\label{eq::rmatrix}
    \braket{A^{a}_{\bar{w}}(z,w)\,A_w^{b} (z',w')}=\frac{C^{ab}}{2i\pi (z-z')}\delta(w-w')\ \ ,
\end{equation}

\noindent where $C$ is the (second) Casimir operator of the Lie algebra (here, $\frac{1}{2}h\otimes h+ e\otimes f + f\otimes e$). Now, multiplying the propagators by the couplings of the the vertices that they relate, we get the full effective action:
\begin{equation}
    S_{eff}=\int_{\Sigma}\mathcal{L}^{\text{chiral}}_{free}+\mathcal{L}_{free}^{\text{anti-chiral}}+\frac{\left(\mathcal{V}^{e}\mathcal{\bar{V}}^{f}+\mathcal{V}^{f}\mathcal{\bar{V}}^{e}+\frac{1}{2}\mathcal{V}^{h}\mathcal{\bar{V}}^{h}\right)T\bar{T}}{2i\pi(z_0-z_1)}\ \ ,
\end{equation}
which explicitly involves a term proportional to $T\bar{T}$. The coefficient $\frac{1}{2i\pi(z_0-z_1)}$, which is here a parameter of the problem, can be taken to be the $\lambda$ deformation parameter of the seed CFT (that we have to keep perturbative to be consistent with our first order approach). On the other hand, the term coming from the Casimir operator and the powers of $w$ plays the role of a metric on $\Sigma$.

\section{$T\bar{T}$ deformation on curved space}\label{sec::TTbar}

In this section we shall explore the possible $T\bar{T}$-deformations that we can obtain using two different groups ($SL_2$ and $SO(2)\ltimes \mathbb{R}^2$) and two different holomorphic 1-forms of the complex manifold ($dz$ and $\frac{dz}{z}$). The first two subsections will be devoted to $SL_2$ (with both choices of 1-forms), the last two will cover the case $SO(2)\ltimes \mathbb{R}^2$.
\subsection{Projective space and AdS space}\label{sec::TTbarCP1}
\subsubsection{$T\bar{T}$ deformation on $\mathbb{CP}^1$}
Consider a 4D Chern-Simons theory on $\mathbb{C}_z\times \mathbb{CP}^{1}_{w}$ symmetry group $SL_2(\mathbb{R})$. Let $(z,w)\in \mathbb{C}\times \mathbb{CP}^1$ be the coordinates of a point in the 4D space, and let the 4D CS theory be coupled to a chiral (resp. antichiral) 2D fermionic CFT living on $\mathbb{CP}^1$ at a specific points $z_0$ (resp $z_1$) of $\mathbb{C}$ using the $T\coloneqq  T_{ww}$ (resp. $\bar{T}\coloneqq \bar{T}_{\bar{w}\bar{w}}$) component of the stress energy tensor of said CFT.

Let us now apply the procedure explained in the previous section to recover from our 4D theory with surface defects a deformed 2D effective field theory. Certainly, we expect to recover our 2D CFT that currently plays the role of surface defect in the 4D theory, together with an interaction term arising from the coupling between the CS gauge field and the stress tensor CFT. We finally hope for the powers of $w$ to arrange into a metric term that the $\lambda T\bar{T}$ term would be integrated against. Let us carry out explicitly this computation.

We will be working to first order in perturbation theory under the gauge fixing given in \ref{sec::4DCST}: consequently, the cubic vertex originally present in the CS action vanishes and we only have the two types of vertices coming from the coupling terms. This gives rise to a limited number of Feynman diagrams that should be all accounted for in order to find the 2D effective action.  As given in Section \ref{sec::4DCST}, the color-stripped propagator is just $\frac{1}{z_{0} -z_{1}}$ and the only possible pairings of gauge fields involving a non-zero propagator are given by the quadratic Casimir operator of $\mathfrak{sl}_2$ :
\begin{equation}\label{eq::casimirsl2}
	C_2(\mathfrak{sl}_2)=e\otimes f+f\otimes e+\frac{1}{2}h\otimes h\ \ ,
\end{equation}
which means that we should consider only three ways to pair up the coupling terms, given by the following Feynman diagrams:

\vspace{0.1cm}
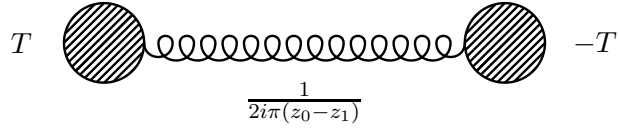
\begin{figure}[H]
\begin{center}
\begin{fmffile}{efCas}  %creates file named three_body to save your diagram
\begin{fmfgraph*}(150,40)    %size of your diagram you can change it as you wish
\fmfleft{e1}              %indicates the particles which will stand on LHS, for sake of simplicity I denoted them as electrons but you can change them as you like
\fmfright{g}

\fmfv{decor.shape=circle, decor.filled=shaded,decor.size=0.2w,label.dist=0.18w}{e1}%indicates the particles will stand on RHS
\fmfv{decor.shape=circle, decor.filled=shaded,decor.size=0.2w,label.dist=0.17w}{g}
\fmflabel{$T$}{e1}         %names which will be shown on your PDF
\fmflabel{$-T$}{g}
\fmf{gluon,label.dist=0.10w,label.side=bottom,label=$\frac{1}{2i\pi(z_0-z_1)}$}{e1,g}       %fermion lines, if you don't want arrows simply write "plain"
%instead of "fermion"
\end{fmfgraph*}
\end{fmffile}
\end{center}

\caption{e-f part of the Casimir}
\label{feynman::interactionD1}
\end{figure}

\vspace{0.1cm}
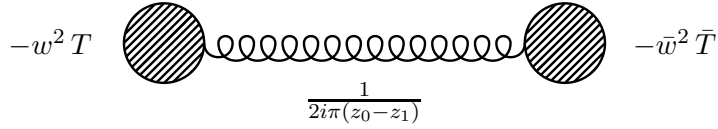
\begin{figure}[H]
\begin{center}
\begin{fmffile}{feCas}  %creates file named three_body to save your diagram
\begin{fmfgraph*}(150,40)    %size of your diagram you can change it as you wish
\fmfleft{e1}              %indicates the particles which will stand on LHS, for sake of simplicity I denoted them as electrons but you can change them as you like
\fmfright{g}

\fmfv{decor.shape=circle, decor.filled=shaded,decor.size=0.2w,label.dist=0.18w}{e1}%indicates the particles will stand on RHS
\fmfv{decor.shape=circle, decor.filled=shaded,decor.size=0.2w,label.dist=0.17w}{g}
\fmflabel{$-w^2\,T$}{e1}         %names which will be shown on your PDF
\fmflabel{$-\bar{w}^2\,\bar{T}$}{g}
\fmf{gluon,label.dist=0.10w,label.side=bottom,label=$\frac{1}{2i\pi(z_0-z_1)}$}{e1,g}       %fermion lines, if you don't want arrows simply write "plain"
%instead of "fermion"
\end{fmfgraph*}
\end{fmffile}
\end{center}

\caption{f-e part of the Casimir}
\label{feynman::interactionD2}
\end{figure}

\vspace{0.1cm}
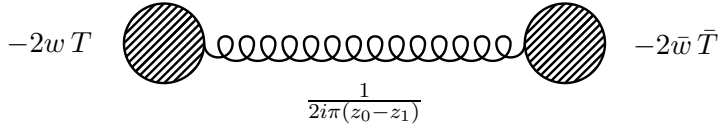
\begin{figure}[H]
\begin{center}
\begin{fmffile}{hhCas}  %creates file named three_body to save your diagram
\begin{fmfgraph*}(150,40)    %size of your diagram you can change it as you wish
\fmfleft{e1}              %indicates the particles which will stand on LHS, for sake of simplicity I denoted them as electrons but you can change them as you like
\fmfright{g}

\fmfv{decor.shape=circle, decor.filled=shaded,decor.size=0.2w,label.dist=0.18w}{e1}%indicates the particles will stand on RHS
\fmfv{decor.shape=circle, decor.filled=shaded,decor.size=0.2w,label.dist=0.17w}{g}
\fmflabel{$-2w\,T$}{e1}         %names which will be shown on your PDF
\fmflabel{$-2\bar{w}\,\bar{T}$}{g}
\fmf{gluon,label.dist=0.10w,label.side=bottom,label=$\frac{1}{2i\pi(z_0-z_1)}$}{e1,g}       %fermion lines, if you don't want arrows simply write "plain"
%instead of "fermion"
\end{fmfgraph*}
\end{fmffile}
\end{center}

\caption{h-h part of the Casimir}
\label{feynman::interactionD3}
\end{figure}

\noindent By summing the contributions of all these graphs, one gets the following effective quartic interaction:
\begin{equation}\label{eq::effquartintCP}
    I_{eff}=\frac{1+|w|^4 + 2*|w|^2}{2i\pi(z_1-z_0)}*T\bar{T}(w)=\frac{(1+|w|^2)^2}{2i\pi(z_1-z_{0})}* T\bar{T}(w)\ \ ,
\end{equation}
where we can recognize at the numerator of the fraction the inverse of the complex component of the Fubini-Study metric for $\mathbb{CP}^1$:
\begin{equation}\label{eq::FSmetric}
    g_{\mathbb{CP}^1}(w,\bar{w})=\frac{1}{(1+|w|^2)^2}\mathrm{d}w\mathrm{d}\bar{w} \implies g_{\mathbb{CP}^1}^{-1}(w,\bar{w})=(1+|w|^2)^2\ \ .
\end{equation}
We can therefore rewrite our 2D effective action, denoting by $\lambda_{01}=\frac{1}{2i\pi(z_1-z_0)}$ the first-order deformation parameter, as:
\begin{equation}\label{eq::finaleffquartintCP1}
    S_{eff}=S^{kin}_{c}+S^{kin}_{ac}+\int_{\mathbb{CP}^1}\lambda_{01}\cdot T\bar{T}(w)\cdot  g^{-1}(w)\ \mathrm{d}w\mathrm{d}\bar{w}\ \ ,
\end{equation}
where the subscripts $c$ and $ac$ refer respectively to chiral and antichiral theories. 
Clearly, we just get a deformation of our theory by a term proportional to $T\bar{T}$ through a coupling constant $\lambda_{01}$ coming from our choice of position of the surface defects in the initial 4D space. 

Let us now take this problem backwards. Had we a 2D chiral theory on $\mathbb{CP}^1$ and should we want to $T\bar{T}$-deform it, we would need to integrate $T\bar{T}$ versus a term that makes dimensional sense. Now $T$ has dimension $(2,0)$ ad $\bar{T}$ has dimensions $(0,2)$, and we want to have an integrand with dimensions $(1,1)$ (like $\mathrm{d}w\mathrm{d}\bar{w}$). We therefore need a quantity with dimensions $(-1,-1)$ to integrate against, and the most natural (and only?) candidate for this definitely seems to be the inverse of the metric.

Thus, from our 4D CS picture with surface defects, we were able to recover a general $T\bar{T}$ deformation of the associated 2D chiral CFT with tunable coupling determined by the initial data of the position of the surface defects in 4D space. This process offers a new point of view on $T\bar{T}$ deformations on curved space, proposing an interpretation thereof in terms of surface operators in 4D Chern-Simons. 
\subsubsection{$T\bar{T}$ deformation on $AdS_2$}

The $T\bar{T}$ deformation of a CFT on $ADS_2$ spacetime can be obtained in a very similar fashion. Indeed, noting that the metric on $AdS$ space differs from that on $\mathbb{CP}^1$ by one minus sign only:
\begin{equation}\label{eq::metricAdS2}
g_{AdS_2}=\frac{\mathrm{d}w\mathrm{d}\bar{w}}{(1-|w|^2)^2}\ \ ,
\end{equation}
it suffices to find another representation of the Virasoro $\mathfrak{sl}_2$ algebra such that the quadratic Casimir yields the same term with an extra minus sign. To that end, one can just modify the Equations \ref{eq::virasorocoeffsCP1} in the following way:

\begin{align}\label{eq::virasorocoeffsCP1p}
    \mathcal{V}^{e}(w)&=i\nonumber \\
    \mathcal{V}^{f}(w)&=-i w^2 \\
    \mathcal{V}^{h}(w)&=-2w\nonumber \ \ ,
\end{align}
and same for the anti-holomorphic part with the couplings for $e$ and $f$ swapped. Following the same steps it is then straightforward to show that

\begin{equation}\label{eq::finaleffquartintAdS}
     S_{eff}=S^{kin}_{c}+S^{kin}_{ac}-\int_{AdS_2}\lambda_{01}\cdot T\bar{T}(w)\cdot  g^{-1}(w)\ \mathrm{d}w\mathrm{d}\bar{w}\ \ .
\end{equation}

\subsection{Deformed sphere endowed with the "sausage metric"}\label{sec::TTbarsausage}
As explained in Section \ref{sec::4DCST}, we can choose several different 1-forms to supplement the 3D CS theory and make it four-dimensional \cite{Costello:2017dso}. In particular, instead of just considering $\mathrm{d}z$ with $z\in \mathbb{C}$, one can take the holomorphic 1-form $\frac{\mathrm{d}z}{z}$ for $z\in \mathbb{C}$. Based on calculations by Costello et al. \cite{Costello:2018gyb}, this changes the shape of the two-point correlation function of the 4D Gauge field. This propagator now takes the form of what Costello et al. introduce as the trigonometric classical $r$-matrix (Section 9.4) in their paper, thereby changing the weight given to each of the previous Feynman diagrams (alternatively, changing the form of the propagators in said diagrams):

\begin{equation}
    r=\frac{z_1}{z_1-z_0}e\otimes f+\frac{z_0}{z_1-z_0}f\otimes e
+\frac{1}{4}\frac{z_1+z_0}{z_1-z_0}\ \ ,\end{equation}
where each term corresponds to the propagator between two terms in the action.

Combining the expression for the $r$-matrix given in \cite{Costello:2018gyb} with the Feynman diagram approach used for the previous cases, we get the following expression after summing the contributions: 

\begin{equation}\label{eq::effquartintSausage}
    I_{eff}(w)=\left(\frac{z_1}{z_1-z_0}*1+\frac{z_0}{z_1-z_0}*|w|^4+\frac{1}{4}\frac{z_1+z_0}{z_1-z_0}*|-2w|^2\right)T\bar{T}(w)\ \ .
\end{equation}
Denoting by $\alpha=\frac{z_1}{z_1-z_0}$ and by $\beta=\frac{z_0}{z_1-z_0}$ (note that $\alpha-\beta=1$), we can rewrite this interaction as follows: 
\begin{equation}\label{eq::effquartintSausage2}
    I_{eff}(w)=\left(\alpha+\beta*|w|^4+(\alpha+\beta)*|w|^2\right)T\bar{T}(w)\ \ .
\end{equation}
By analogy with our previous calculation, this would be the $T\bar{T}$ deformation of a CFT on a space with metric proportional to

\begin{equation}\label{eq::metricneededsausage}
g(w)\propto \frac{\frac{1}{\alpha}\mathrm{d}w\mathrm{d}\bar{w}}{1+\frac{\beta}{\alpha}*|w|^4+(1+\frac{\beta}{\alpha})*|w|^2}\ \ .
\end{equation}
$\frac{\beta}{\alpha}=\frac{z_1}{z_0}$, and we know that the 4D theory is $z$-translation invariant and thus only depends on $(z_0-z_1)$.Consequently, given $z_1-z_0$, we can demand that $\frac{z_1}{z_0}$ be finite non-zero and real by moving $z_0$ freely without loss of generality.

Provided $\frac{\beta}{\alpha}$ is real (which we just made sure of), one can consider the scalar dilation 
\begin{align}\label{eq::dilationsausage}
w'&=\left(\frac{\beta}{\alpha}\right)^{\frac{1}{4}}*w \nonumber\\
\bar{w}'&=\left(\frac{\beta}{\alpha}\right)^{\frac{1}{4}}*\bar{w}\ \ ,
\end{align}
which allows us to rewrite the expected metric as follows:

\begin{equation}\label{eq::metricneededsausage2}
g(w')\propto \frac{\frac{1}{\sqrt{\beta*\alpha}}\mathrm{d}w'\mathrm{d}\bar{w}'}{1+|w'|^4+(\sqrt{\frac{\beta}{\alpha}}+\sqrt{\frac{\alpha}{\beta}})*|w'|^2}\ \ .
\end{equation}
Now introducing the variable
\begin{equation}\label{eq::tvariablesausage}
    t=-\frac{1}{8}\ln\left(\frac{\beta}{\alpha}\right)\implies e^{4t}=\sqrt{\frac{\alpha}{\beta}}\ \ ,
\end{equation}
and dropping the primes, we can finally rewrite the metric of our unknown space $g(w)$ as the so-called "Sausage metric" \cite{Bykov:2020llx,Lambert:2012tq} with parameter $t$ as introduced by Lambert et al.:

\begin{equation}\label{eq::sausagemetric}
g(w)\propto \frac{2\sinh(4t)\ \mathrm{d}w\mathrm{d}\bar{w}}{1+|w|^4+2\cosh(4t)|w|^2}\ \ .
\end{equation}
Note that to get this, we used that $2\sinh(4t)=\sqrt{\frac{\alpha}{\beta}}-\sqrt{\frac{\beta}{\alpha}}=\frac{\alpha-\beta}{\sqrt{\alpha\beta}}=\frac{1}{\sqrt{\beta\alpha}}$ by definition of $\alpha$ and $\beta$.

This shows that the 2D action obtained from the 4D CS theory with holomorphic 1-form $\frac{\mathrm{d}z}{z}$ and gauge group $SL_2$ in the presence of fermionic surface defects on a deformed sphere endowed with the Sausage metric yields an effective action that turns out to be the $T\bar{T}$ deformation of the CFT on that deformed sphere:

\begin{equation}\label{eq::finaleffquartintsausage}
    S_{eff}=S^{kin}_{c}+S^{kin}_{ac}+\int_{S^2}T\bar{T}(w)\cdot  g^{-1}(w)\ \mathrm{d}w\mathrm{d}\bar{w}\ \ .
\end{equation}

\subsection{$T\bar{T}$ deformation on a space with a cigar metric}\label{sec::cigar}
So far, we have used only $SL(2,\mathbb{R})$ symmetry for our 4D CS theory supplemented by its Virasoro counterpart to have invariant coupling terms with 2D CFTs. But we could also consider other symmetry groups $G$, as long as we can engineer coupling terms with the energy-momentum tensors that also present such a symmetry and thus allow to consider a diagonal subgroup of $G\times G$. In particular, the Lie algebra isometries of the plane $SO(2)\ltimes \mathbb{R}^2$ has a similar representation to that of $SL(2,\mathbb{R})$, with only one -sizeable- difference in the commutators. Namely, $Lie\left(SO(2)\ltimes \mathbb{R}^2\right)$ is generated by $e,f$ and $h$ such that:
\begin{align}
    [h,e]&=2e\nonumber \\
    [h,f]&=-2f\\
    [e,f]&=0\ \ ,\nonumber
\end{align}
from which we could assert that $sl(2)$ is but a deformation of this algebra to yield a non-trivial $[e,f]$ Lie bracket.
If we assign this symmetry to our Gauge field $A$, we need to adjust the couplings for it to be also respected by the operator $\oint T*$, which is easily done. It suffices indeed to impose $\mathcal{V}^{f}=0$ and $\bar{\mathcal{V}}^{e} =0$. Now, the more subtle issue is to determine the new shape of the propagator when the symmetry group is the isometries of the plane, that is, find a non-degenerate bilinear 2-form (the Killing form is trivially degenerate) from which to extract the quadratic Casimir. Calling this bilinear form $q$, we can construct it by hand by specifying its value on all of the pairs of generators, and check that it respects the Lie algebra structure. We impose the following equations:

\begin{align}
    q(e,f)&=1\nonumber \\
    q(h,h)&=1\\
    q(e,h)&=q(f,h)=0\ \ ,\nonumber \\
\end{align}
extended by symmetry and $\mathbb{R}$-linearity to a symmetric bilinear map. It is non-degenerate by construction, and we can check that
\begin{equation}
    q(x,[y,z])=q([x,y],z)
\end{equation}
very easily.

This construction leads to a Casimir that turns out to be identical to that of $SL(2)$. Combining this finding with our knowledge of the new couplings $(\mathcal{V}^{a},\bar{\mathcal{V}}^{a})$ and using the same Feynman diagrams as in Section \ref{sec::TTbarCP1}, we now get:
\begin{equation}
    S_{eff}=S_{kin}+\frac{1}{z_0-z_1}\int_{\mathcal{C}} T\bar{T}(1+|w|^2)\mathrm{d}^2 w\ \,
\end{equation}
which suggests that we would be able to $T\bar{T}$ deform a space $\mathcal{C}$ that is endowed with the metric 
\begin{equation}
    g(w)=\frac{\mathrm{d}w \mathrm{d}\bar{w}}{1+|w|^2}\ \ ,
\end{equation}
which is precisely the so-called cigar metric.

\subsection{$T\bar{T}$ deformation on a deformed cigar}

Once again, we can wonder if we could recover the $T\bar{T}$-deformation of a different kind of curved space using a different holomorphic 1-form in our initial action: $\frac{dz}{z}$. Following the same steps as in Section \ref{sec::TTbarsausage} (and the same notation) but using this time the commutation relations and couplings introduced in Section \ref{sec::cigar}, we get an expression of the metric for the deformed space that reads
\begin{equation}
    g(w')=-\frac{2\sinh(4t)\ \mathrm{d}w'\mathrm{d}\bar{w}'}{1+2\cosh(4t)\ |w'|^2}\ \ ,
\end{equation}
which means that we recover the first-order $T\bar{T}$-deformation of a space endowed with a slightly deformed version of the cigar metric. Although this deformation does not differ much from the one derived in the previous subsection, it is worth noting that one can easily tune the deformation of the cigar metric using the parameters at our disposal ($z_0$ and $z_1$). 

\section{Conclusion}

In this paper we endeavored to interpret curved space $T\bar{T}$-deformations of 2D chiral + antichiral CFTs to first order in perturbation theory as coming from a pair of chiral and antichiral surface CFTs in 4D gauge theories with a suitable coupling scheme. This led to a fairly straightforward 4D interpretation of the deformation parameter as the inverse of the 2D ``holomorphic distance'' ($z_1-z_0$) between the two CFT planes. Moreover, by slightly changing the gauge group of our 4D theory and playing with different holomorphic 1-forms of our complex manifold, we could recover $T\bar{T}$-deformations on spaces of various curved geometries, including projective spaces, and spaces endowed with metrics such as the so-called cigar and sausage metrics. 

There are many future directions to explore. Our method only allowed us to retrieve first-order deformations of the action. It would be interesting to compute the effective action at higher order (which is significantly more involved and has not been attempted here) and check that the result matches the expected $T\bar{T}$-deformed theory. Moreover, explicit parallels between $T\bar{T}$-deformations and 3D gravity have previously been drawn, and it is well known that 3D gravity is closely related to 3D Chern-Simons theory. This motivates one to wonder if compactifying 4D Chern-Simons theory along the lines described in Section \ref{sec::gravity} could lead to the same gravity theories as the one we would obtain by using the direct bridge between $T\bar{T}$-deformations and gravity.
\\

\section*{Acknowledgements}

I would like to thank Kevin Costello, who suggested this project and helped and guided me through it. I am also very grateful to my advisor Tudor Dimofte for guidance in preparation for this manuscript. Lastly, I am thankful to Leonel Quinta Queimada for many useful discussions. 
\bibliographystyle{JHEP}
\bibliography{references}

\providecommand{\href}[2]{#2}\begingroup\raggedright\begin{thebibliography}{10}

\bibitem{Mazenc:2019cfg}
E.A.~Mazenc, V.~Shyam and R.M.~Soni, \emph{{A $T \bar{T}$ Deformation for
  Curved Spacetimes from 3d Gravity}},
  \href{https://arxiv.org/abs/1912.09179}{{\ttfamily 1912.09179}}.

\bibitem{Cardy:2018sdv}
J.~Cardy, \emph{{The $ T\overline{T} $ deformation of quantum field theory as
  random geometry}}, \href{https://doi.org/10.1007/JHEP10(2018)186}{\emph{JHEP}
  {\bfseries 10} (2018) 186}
  [\href{https://arxiv.org/abs/1801.06895}{{\ttfamily 1801.06895}}].

\bibitem{Smirnov:2016lqw}
F.A.~Smirnov and A.B.~Zamolodchikov, \emph{{On space of integrable quantum
  field theories}},
  \href{https://doi.org/10.1016/j.nuclphysb.2016.12.014}{\emph{Nucl. Phys. B}
  {\bfseries 915} (2017) 363}
  [\href{https://arxiv.org/abs/1608.05499}{{\ttfamily 1608.05499}}].

\bibitem{Cavaglia:2016oda}
A.~Cavagli\`a, S.~Negro, I.M.~Sz\'ecs\'enyi and R.~Tateo, \emph{{$T
  \bar{T}$-deformed 2D Quantum Field Theories}},
  \href{https://doi.org/10.1007/JHEP10(2016)112}{\emph{JHEP} {\bfseries 10}
  (2016) 112} [\href{https://arxiv.org/abs/1608.05534}{{\ttfamily
  1608.05534}}].

\bibitem{Jiang:2019epa}
Y.~Jiang, \emph{{A pedagogical review on solvable irrelevant deformations of 2D
  quantum field theory}},
  \href{https://doi.org/10.1088/1572-9494/abe4c9}{\emph{Commun. Theor. Phys.}
  {\bfseries 73} (2021) 057201}
  [\href{https://arxiv.org/abs/1904.13376}{{\ttfamily 1904.13376}}].

\bibitem{Brennan:2020dkw}
T.D.~Brennan, C.~Ferko, E.~Martinec and S.~Sethi, \emph{{Defining the $T
  \overline{T}$ Deformation on $\mathrm{AdS}_2$}},
  \href{https://arxiv.org/abs/2005.00431}{{\ttfamily 2005.00431}}.

\bibitem{Dubovsky:2017cnj}
S.~Dubovsky, V.~Gorbenko and M.~Mirbabayi, \emph{{Asymptotic fragility, near
  AdS$_{2}$ holography and $ T\overline{T} $}},
  \href{https://doi.org/10.1007/JHEP09(2017)136}{\emph{JHEP} {\bfseries 09}
  (2017) 136} [\href{https://arxiv.org/abs/1706.06604}{{\ttfamily
  1706.06604}}].

\bibitem{Gross:2019ach}
D.J.~Gross, J.~Kruthoff, A.~Rolph and E.~Shaghoulian, \emph{{$T\overline{T}$ in
  AdS$_2$ and Quantum Mechanics}},
  \href{https://doi.org/10.1103/PhysRevD.101.026011}{\emph{Phys. Rev. D}
  {\bfseries 101} (2020) 026011}
  [\href{https://arxiv.org/abs/1907.04873}{{\ttfamily 1907.04873}}].

\bibitem{Tolley:2019nmm}
A.J.~Tolley, \emph{{$ T\overline{T} $ deformations, massive gravity and
  non-critical strings}},
  \href{https://doi.org/10.1007/JHEP06(2020)050}{\emph{JHEP} {\bfseries 06}
  (2020) 050} [\href{https://arxiv.org/abs/1911.06142}{{\ttfamily
  1911.06142}}].

\bibitem{Witten:1988hc}
E.~Witten, \emph{{(2+1)-Dimensional Gravity as an Exactly Soluble System}},
  \href{https://doi.org/10.1016/0550-3213(88)90143-5}{\emph{Nucl. Phys. B}
  {\bfseries 311} (1988) 46}.

\bibitem{Costello:2017dso}
K.~Costello, E.~Witten and M.~Yamazaki, \emph{{Gauge Theory and Integrability,
  I}}, \href{https://doi.org/10.4310/ICCM.2018.v6.n1.a6}{\emph{ICCM Not.}
  {\bfseries 06} (2018) 46} [\href{https://arxiv.org/abs/1709.09993}{{\ttfamily
  1709.09993}}].

\bibitem{Costello:2019tri}
K.~Costello and M.~Yamazaki, \emph{{Gauge Theory And Integrability, III}},
  \href{https://arxiv.org/abs/1908.02289}{{\ttfamily 1908.02289}}.

\bibitem{Dubovsky:2018bmo}
S.~Dubovsky, V.~Gorbenko and G.~Hern\'andez-Chifflet, \emph{{$ T\overline{T} $
  partition function from topological gravity}},
  \href{https://doi.org/10.1007/JHEP09(2018)158}{\emph{JHEP} {\bfseries 09}
  (2018) 158} [\href{https://arxiv.org/abs/1805.07386}{{\ttfamily
  1805.07386}}].

\bibitem{Iliesiu:2020zld}
L.V.~Iliesiu, J.~Kruthoff, G.J.~Turiaci and H.~Verlinde, \emph{{JT gravity at
  finite cutoff}},
  \href{https://doi.org/10.21468/SciPostPhys.9.2.023}{\emph{SciPost Phys.}
  {\bfseries 9} (2020) 023} [\href{https://arxiv.org/abs/2004.07242}{{\ttfamily
  2004.07242}}].

\bibitem{Isler:1989hq}
K.~Isler and C.A.~Trugenberger, \emph{{A Gauge Theory of Two-dimensional
  Quantum Gravity}},
  \href{https://doi.org/10.1103/PhysRevLett.63.834}{\emph{Phys. Rev. Lett.}
  {\bfseries 63} (1989) 834}.

\bibitem{Chamseddine:1989yz}
A.H.~Chamseddine and D.~Wyler, \emph{{Gauge Theory of Topological Gravity in
  (1+1)-Dimensions}},
  \href{https://doi.org/10.1016/0370-2693(89)90528-5}{\emph{Phys. Lett. B}
  {\bfseries 228} (1989) 75}.

\bibitem{Costello:2018gyb}
K.~Costello, E.~Witten and M.~Yamazaki, \emph{{Gauge Theory and Integrability,
  II}}, \href{https://doi.org/10.4310/ICCM.2018.v6.n1.a7}{\emph{ICCM Not.}
  {\bfseries 06} (2018) 120}
  [\href{https://arxiv.org/abs/1802.01579}{{\ttfamily 1802.01579}}].

\bibitem{Bykov:2020llx}
D.~Bykov and D.~Lust, \emph{{Deformed $\sigma$-models, Ricci flow and Toda
  field theories}},
  \href{https://doi.org/10.1007/s11005-021-01484-0}{\emph{Lett. Math. Phys.}
  {\bfseries 111} (2021) 150}
  [\href{https://arxiv.org/abs/2005.01812}{{\ttfamily 2005.01812}}].

\bibitem{Lambert:2012tq}
C.~Lambert and V.~Suneeta, \emph{{Stability analysis of the Witten black hole
  (cigar soliton) under world-sheet RG flow}},
  \href{https://doi.org/10.1103/PhysRevD.86.084041}{\emph{Phys. Rev. D}
  {\bfseries 86} (2012) 084041}
  [\href{https://arxiv.org/abs/1205.3043}{{\ttfamily 1205.3043}}].

\end{thebibliography}\endgroup
\end{document}